\documentclass[preprints,article,accept,oneauthor,pdftex]{mdpi_mod} 

\Title{Use of Redshifts as Evidence of Dark Energy}

\TitleCitation{Use of Redshifts as Evidence of Dark Energy}

\Author{Jan Stenflo $^{1,2}$}

\address{%
$^{1}$ \quad Institute for Particle Physics and Astrophysics, ETH
(Eidgen\"ossische Technische Hochschule) 
Zurich, CH-8093 Zurich, Switzerland; stenflo@astro.phys.ethz.ch\\ 
$^{2}$ \quad 
Istituto Ricerche Solari ``Aldo e Cele Dacc\`o'' (IRSOL), 
Universit\`a  della Svizzera Italiana, via Patocchi 57, CH-6605
Locarno, Switzerland} 

\newcommand{\aj}{Astron.~J.} 
\newcommand{\apj}{Astrophys.~J.} 
\newcommand{\apjl}{Astrophys.~J.~Lett.} 
 
\newcommand{\aap}{Astron.~Astrophys.} 
\newcommand{\araa}{Ann.~Rev.~Astron.~Astrophys.} 
\newcommand{\aapr}{Astron.~Astrophys.~Rev.} 
\newcommand{\mnras}{Mon.~Not.~R.~Astron.~Soc.} 
 
\newcommand{\prl}{Phys.~Rev.~Lett.} 
\newcommand{\jcap}{J.~Cosmol. Astropart.~Phys.} 

\abstract{
The large-scale dynamics of the universe is 
  generally described in terms of the time-dependent scale factor
  $a(t)$. To make contact with observational data, the $a(t)$ function
  needs to be related to the observable $z(r)$ function, redshift
   versus distance.  
   Model fitting of data has shown that the
  equation that governs $z(r)$ needs to contain a constant term, which
  has been identified as Einstein's cosmological constant. Here, it is
  shown that the required constant term is not a cosmological constant
  but is due to an overlooked
  geometric difference between proper
  time $t$ and look-back time $t_{\rm lb}$ along lines of sight, which
  fan out isotropically in all directions of the 3D (3-dimensional)
  space that constitutes the observable universe. The constant term is needed to
  satisfy the requirement of spatial isotropy in the local limit. Its
  magnitude is independent of the epoch in which the observer
  lives and agrees with the value found by model fitting of
  observational data. Two of the observational
  consequences of this explanation are examined: an increase in the
  age of the universe from 13.8~Gyr 
to 15.4~Gyr, and a resolution of the
  $H_0$ tension, which restores consistency to cosmological theory. 
 }


\begin{document}

\section{Introduction}\label{sec:intro}
The physical nature of the cosmological constant $\Lambda$ that was 
introduced by \citet{stenflo-einstein17} a century ago has remained an 
enigma. It was unexpectedly found necessary to reintroduce $\Lambda$
in 1998 as a fitting parameter to allow for modeling of redshift 
$z$  versus distance $r$ in terms of the FLRW 
(Friedmann--Lema\^itre--Robertson--Walker) framework for observations 
of supernovae type Ia as standard
candles~\cite{stenflo-riessetal1998a,stenflo-perlmutteretal1999a}. 
If the $\Lambda$ term is placed on the 
right-hand side of the Einstein field equation and considered as a 
physical field that is a component $\rho_\Lambda$ (commonly referred 
to as ``dark energy'') of the 
energy-momentum tensor, then the observed sign and magnitude of 
this field represents 
a repulsive gravity force that permeates all of 
space without any significant spatial structuring, driving an 
accelerated cosmic expansion; 
see~\cite{stenflo-Frieman_review2008,stenflo_durrer2011,stenflo-binetruy2013,
stenflo-joyce_review2016,stenflo-amendola2018,stenflo-sabulsky2019}.
Furthermore, as~the 
magnitude of $\rho_\Lambda$ is found to be of the same order as the 
matter density $\rho_M$, although~$\rho_M$ varies with redshift as 
$(1+z)^3$ while $\rho_\Lambda$ is independent of $z$, the~present 
epoch seems to be singled out as 
special. Such a ``cosmic coincidence'' violates 
the Copernican principle, which states that we are not 
privileged~observers. 

While the supernovae observations that were reported in 1998
represented the remarkable discovery of an unexpected property of the
redshift-distance relation $z(r)$, the~interpretation of the 
redshift data in terms of an accelerated cosmic expansion as driven by some
``dark energy'' depends on a theoretical model for the relation
between the observable $z(r)$ function and the $a(t)$ function that
describes the dynamics and evolution of the universe in terms of scale
factor $a$  versus time $t$. As~the time dependence 
of the scale factor is not directly observable, it is inferred from a
static historical record of a 
sequence of past discrete events like in archaeology. In~the case of
cosmology, the static timeline of past 
historical events is accessed by looking out in
space. Because of the finite speed of light, distance from the observer
and look-back time are equivalent coordinates. 
 When cosmic distances are measured with the help of the astrophysical ``distance ladder''---which makes use of ``standard candles''---in~particular supernovae, 
 the corresponding look-back~times are also obtained. 

This ``cosmic archeology'' can allow us to infer the
 dynamical properties of the universe, but~only if it is known how to
 relate the observable, ``archeological'' $z(r)$ function to the
dynamic, non-observable $a(t)$ function. In~the past, the conversion
from $a(t)$ to $z(r)$ has been taken care of in a straightforward way:
in the FLRW equation that governs the $a(t)$ 
function, one replaces $a$ with $z$ via the relation $a=1/(1+z)$ and
identifies the expansion rate ${\dot a}/a$ with the Hubble constant
$H\equiv {\rm d}z/{\rm d}r$, where the dot denotes the time derivative.
The~variation $a(t_{\rm lb})$ with
distance $r$ or look-back time $t_{\rm lb} $ is obtained from the FLRW
solution for the $a(t)$ function. 

This procedure of standard cosmology has
not been seen as a contentious issue. However, it overlooks a
fundamental difference between dynamic, proper time $t$ and look-back
time $t_{\rm lb}$. The~function $a(t_{\rm lb})$ is isotropic with
spherical symmetry in a static 3D (3-dimensional) 
space, the~observable universe. The~symmetry is broken when a redshift is
determined, because~the observation implies the selection of the
particular line of sight that connects the observer with the observed
object. In~the local limit, however, the~requirement of spatial
isotropy must always be~satisfied.

It is shown in the present paper that when this isotropy requirement 
is accounted for, the equation that governs the $z(r)$ 
 relation acquires an additional constant term $\Omega_X =2/3$, which formally
appears in the same way as a cosmological constant $\Omega_\Lambda$
in standard cosmology. $\Omega_X$ is not, however, a cosmological
constant, because~it 
does not appear in the FLRW equation that governs the $a(t)$
function. Its 2/3 value does not depend on the epoch of the
observer. There is no ``cosmic coincidence'' problem. Other
implications of the theory include an increase of the age of the universe
from 13.8~Gyr to 15.4~Gyr and a resolution of the $H_0$ tension.

\section{Relation Between Redshift, Time, and~Distance}\label{sec:redshiftexpress}
The formal relation between redshift $z$ and scale factor $a$ is seemingly simple,
\begin{equation}
1+z\,=\, a_0/a 
\label{eq:a0overa}\end{equation}
(with normalization factor $a_0$), but the relation \eqref{eq:a0overa} may 
 be a source of confused and incorrect interpretations if
one is not clear enough about the coordinate frames that are~used.

An example of an incorrect interpretation is when one considers the
time dependence of this expression. According to a basic FLRW
assumption, the~scale factor is a function of
time $t$ in co-moving frames, but~it does 
not vary with space in 3D hyperplanes that are orthogonal to the time
axis. The normalization parameter $a_0$ is by convention set to
unity at the location of the observer. From~this, one may conclude 
that the redshift is time dependent, as given by $z(t)=-1+1/a(t)$, and
therefore~that $-{\dot z}(t_{\rm obs})=({\dot a}/a)_{\rm obs}$,
the~expansion rate at the location of the~observer. 

It immediately becomes understandable
 that this conclusion is completely wrong, 
when recalling that the spacetime location of the observer defines the
zero point of the 
redshift scale. Therefore, the redshift with all
 its temporal gradients, to~infinite order, becomes zero at the
 location of the observer. Thus, $z_{\rm obs} ={\dot z}_{\rm obs} ={\ddot z}_{\rm obs}
=0$ forever. ${\dot z}_{\rm obs}$ can never equal $-({\dot a}/a)_{\rm
  obs}$. 

The cause for this confusion is the neglect of the time
dependence of $a_0 \equiv a(t_{\rm obs})$. At~the location of the observer,
the time dependence of the numerator and denominator of
Equation~(\ref{eq:a0overa}), ${\dot a}_0$ and ${\dot a}$, become
identical and compensate each 
other, such that the time dependence of $z$ becomes zero, as~required
for consistency with the definition of~redshifts.

 The goal here is to derive the redshift-distance function $z(r)$ from the
$a(t)$ function. The~$r$ dependence of $z$ implies according to
Equation~(\ref{eq:a0overa}) that
\begin{equation}
1+z(r)\,=\, 1/a(r). 
\label{eq:zardep}\end{equation}
However, distance $r$ is not a coordinate in spatial hyperplanes
that are orthogonal to the time axis, because~the scale factor does
not vary within such hyperplanes. Instead, $r$ is a 
coordinate along lines of sight in the 3D hypersurface of
the past light~cone.

Since lines of sight are geodesics of light rays, which are curved
in normal spacetime representations because of the nonlinear cosmic
expansion, it is convenient to use conformal time $\eta$
instead of proper time $t$, as~defined by
\begin{equation}
{\rm d}\eta\equiv {\rm d}t/a. 
\label{eq:conftime}\end{equation}
The special advantage with conformal time is that the nonlinear
expansion is compensated for in a way that makes light rays and
lines of sight straight lines in a spacetime~representation. 

Look-back time $t_{\rm lb}$ is the temporal separation between the
observer and the observed object (e.g., a galaxy):
\begin{equation}
t_{\rm lb}\,\equiv\,t_{\rm obs}\,-\,t\,, 
\label{eq:tlb}\end{equation}
where $t_{\rm obs}$ and $t$ are the ages of the universe at the
spacetime location of the observer and the observed object,
respectively.

In terms of conformal time, 
look-back time $\eta_{\rm lb}$ becomes equivalent to coordinate distance $r$ when
using units, for~which the speed of light $c$ is unity (as 
 used throughout the present paper). Thus,
\begin{equation}
r\equiv \eta_{\rm lb}=\eta_0-\eta,
\label{eq:xlookback}\end{equation}
where $\eta_0$ represents the present conformal age of the universe. 
It allows for a more explicit expression for the redshift:
\begin{equation}
1+z(r,\eta_0)\,=\,\frac{a(\eta_0)}{a(\eta_0 -r)}. 
\label{eq:zaeta0}\end{equation}

If the time dependence $\eta_0$ of the observer is accounted for,
which is the case when the past light cone is treated as a 4D object, then 
the time derivative of $z$ is zero at the observer ($r=0$), as~required by the definition of the redshifts. However, the~distance
derivative $\partial z/\partial r$, which represents the original spatial
definition of the Hubble constant, is non-zero, because~$r$ appears
only in the denominator but not in the numerator, in~contrast to the
time~coordinate.

\begin{figure}[H]
\includegraphics[scale=0.4]{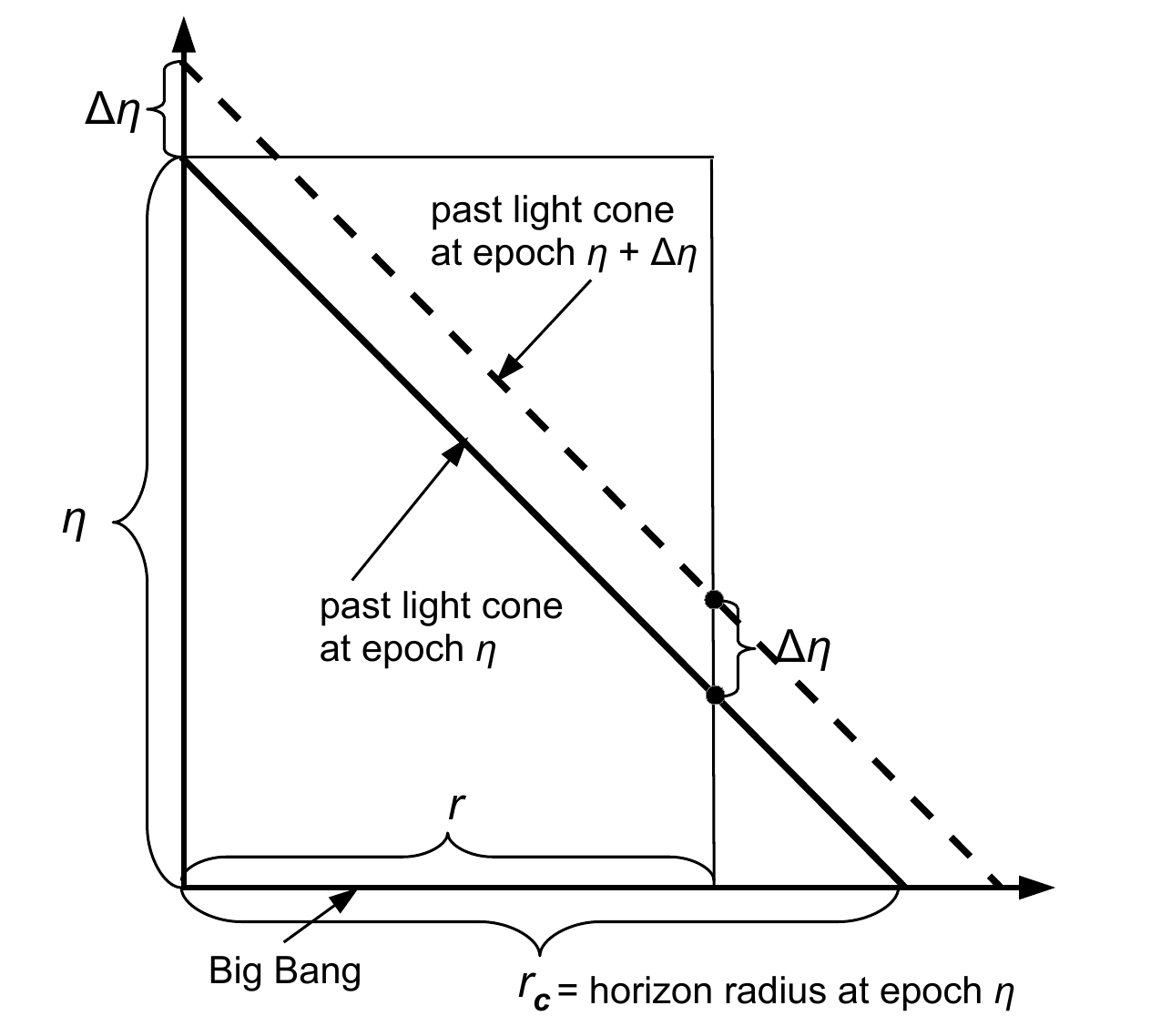}
\caption{Illustration of the past light cone as a time-dependent object in a conformal $\eta ,r$ representation. The~observer moves along the 
vertical time axis together with the tip of the past light cone. An~observed galaxy at distance $r$ also moves along the vertical time direction. Distance $r$ does not change with time because the galaxies in 
 Friedmann--Lema\^itre--Robertson--Walker cosmology are assumed to be
 comoving with the spatial grid. The~radius $r_c$ of the horizon
 equals the conformal age $\eta$ of the universe if one uses  
units where the speed of light $c=1$. 
\label{fig:fig1}}
\end{figure}

The conformal spacetime diagram of Figure~\ref{fig:fig1} illustrates
how the evolution of the past light cone affects the redshift
observations. Redshifts $z$ are only defined along light rays that
represent lines of sight. They vary from zero at the tip of the cone,
where the observer is located, to~infinity at the bottom of the cone,
at the Big Bang, at~a distance $r_c$ (the horizon radius) from the~observer. 

A basic assumption of the FLRW interpretational framework is that all
observed objects are co-moving, at~rest with respect to the
expanding spatial grid. The~coordinate distances $r$ and conformal look-back
times $\eta_{\rm lb}$ are therefore 
time independent (in contrast to proper
distances $ar$). It then follows from  Equation~(\ref{eq:xlookback})
that ${\rm d}\eta_{\rm obs} ={\rm 
  d}\eta$. Conformal time ``flows'' at the same rate for the 
observer as for the observed galaxy, 
independent of its distance. In~contrast, proper time
$t$ flows at different rates because~the scale factor varies with
distance. With~conformal time, the scale factor has been divided
out to~make the relation between distance and conformal time linear
so that light rays become straight~lines.
Only when one formally halts the flow of time for the observer by
treating $\eta_0$ as a constant can one obtain a representation in which 
distance varies with time. Then, as~follows from 
Equation~(\ref{eq:xlookback}), ${\rm d}\eta= -{\rm
  d}\eta_{\rm lb} = -{\rm d}r$. In~this
case, time does not ``flow'' but is a static coordinate along a
historical record of past events. This
description represents a 3D subspace, the~part of 4D spacetime
that is accessible 
to the observer at the moment of observation. It is the subspace that is referred
to as the ``observable universe''. As~the redshift is then only a function of the single
distance parameter $r$ or $\eta_{\rm lb}$, the~partial derivatives can be
replaced by total derivatives: ${\rm d}z/{\rm d}r$ or ${\rm d}z/{\rm
  d}\eta_{\rm lb}$.

\section{Geometric Aspects of Look-Back Time, Spatial Isotropy, and~  Implications for \boldmath$z(r)$}\label{sec:hubble} 
When making the connection between the dynamic $a(t)$ function that
represents 4D spacetime and the
$z(r)$ and $a(r)$ functions that represent the static 3D observable universe, 
different geometric aspects enter 
in a profound way. At~first glance,
$a(r)=a(\eta_{\rm lb})$ according to Equation~(\ref{eq:xlookback}) may
seem to imply that one can simply use the $a(\eta)$
solution for $a(r)$ that follows from the dynamic $a(t)$ function,
and this then immediately solves the problem. While this is what
standard cosmology does, it overlooks the geometric issues
that become relevant when connecting a 1D function to a function that is isotropic
in a 3D vector~space. 

While $t$ and $\eta$ are 1D coordinates, look-back time $t_{\rm lb}$ or
$\eta_{\rm lb}$ are radial coordinates that run along the lines of
sight that fan out in 3D space from the central point, the~location of
the observer, where $r$ and $t_{\rm lb}$ are zero. Although~the $a(\eta_{\rm lb})$
function is identical to the 1D $a(\eta)$ function along each 1D line of
sight, there are infinite viewing directions. As~each line of sight needs
to be defined by three mutually independent parameters, 
there are three spatial degrees of~freedom. 

$a(r)$ is an isotropic function in 
the 3D space that defines the observable universe. The~
act of observation is, however, not isotropic. The~observation of a
given galaxy singles out a certain viewing direction, which
breaks the spatial symmetry. While different galaxies correspond to
different viewing 
directions, each single redshift measurement breaks the~isotropy.

After these introductory remarks, let us see how these concepts play
out for the explicit derivation of the Hubble constant and the
expression for the $z(r)$ function. 
The starting point is the FLRW equation for the
$a(t)$ function with the assumption of spatial flatness and no
cosmological constant:
\begin{equation}
\left(\frac{\dot a}{a}\right)^{\!\!2}\,=\frac{8\pi\, G}{3}\,[\,
\rho_{M_0}(a_0/a)^3\,+\,\rho_{R_0}(a_0/a)^4\,], 
\label{eq:flrwaexpans}\end{equation}
 where subscript 0 refers to the location of the observer, and $G$ is
 the gravitational constant.  

The squared expansion rate that describes the evolution of the
universe has its sources in the mass--energy densities $\rho_{M,\,R}$ of matter and
radiation. Since there is no cosmological constant, the universe is
decelerating.

The first step in making the connection between the time and distance
representations is to express the expansion rate in terms of conformal
time, because~distance and conformal look-back time are equivalent
coordinates according to
Equation~(\ref{eq:xlookback}): 
\begin{equation}
\left(\frac{\dot a}{a}\right)^{\!\!2}\,=\,\left(\frac{{\rm
      d}a}{a\,{\rm
      d}t}\right)^{\!\!2}\,=\,\left(\frac{1}{a^2}\,\frac{{\rm d}a}{{\rm
      d}\eta}\right)^{\!\!2}.  
\label{eq:dadteta2}\end{equation}
In Equation~(\ref{eq:dadteta2}), the time coordinates $t$ and $\eta$
are orthogonal to the spatial directions and represent the time in
co-moving coordinate~frames. 

The difference between co-moving time $\eta$ and its look-back version
$\eta_{\rm lb}$ is, in~terms of a spacetime diagram, that $\eta$ is a 1D
coordinate along the vertical time axis, while $\eta_{\rm lb}$ is a
coordinate along lines that represent light rays at $45^\circ$ to the
time axis, fanning out isotropically in all 3D spatial~directions. 

When  $\eta$ is constrained to run along light rays,  we have to attach
three index labels, representing the three orthogonal spatial
directions ($x,y,z$). Since the observable universe can be treated as
a vector space with the observer at its center, it is convenient to
introduce a Cartesian $x,y,z$ coordinate system 
with its origin at the location of the observer, where the distance
parameter $r$ is~zero. 

Let us for convenience introduce notation $H_x$ for the $\eta$
gradient of the scale factor along a light ray in the $x$ 
direction, defined as follows:
\begin{equation}
 H_x^2\,\equiv\, \, \left(\frac{1}{a^2}\, \frac{{\rm d}a}{{\rm
       d}\eta}\right)^{\!\!2}_{\!\!x}\,=\left(\frac{\dot a}{a}\right)^{\!\!2}.  
    \label{eq:hx2defin}\end{equation}
 The second equality in Equation~(\ref{eq:hx2defin})
  follows from Equation~(\ref{eq:dadteta2}), because~the universe is assumed to be
  intrinsically isotropic with the same variation of the scale factor
  with $t$ and $\eta$ along any light ray, regardless of its
  direction. It is not valid for an anisotropic scale~factor.

Note that the vector components $H_{x,y,z}$ have been defined as
gradients of the scale factor in different spatial directions along
the past light cone, not in terms of redshift gradients. The~notation
that is used 
should at this stage of the derivation not be confused with the $H$
that is the standard notation for the Hubble constant. There
is a fundamental conceptual difference between redshift and scale
factor, in~spite of the seemingly 
simple Equation~(\ref{eq:zardep}) that relates
the two. Scale factor is an observer-independent scalar quantity of 4D
spacetime, which at any location in spacetime can possess gradients in
all directions, e.g.,~in time or in any of the three spatial directions
along a light cone. In~contrast, redshifts and their spatial gradients
only have meaning as scalar quantities along a line of sight that
 emanates from an observer. The~problem that is dealt with here
 is to relate the 3D components of the scale factor gradient to the
relevant 1D redshift gradient in a way that uses boundary conditions
that preserve the intrinsic spatial symmetry of the scale factor. Only
towards the end of this derivation, just before Equation~(\ref{eq:h2omegas}) below,
one can identify the scalar that represents the observable
Hubble parameter. 
  
The gradients in each spatial direction represent the spatial
components of a vector, whose squared radial magnitude is the sum of
the squared coordinate components:
\begin{equation}
H^2\,\equiv H_x^2\,+\, H_y^2\,+\, H_z^2.  
\label{eq:h2defini}\end{equation}

At the center of the spherical coordinate
system, where $r=0$, there is directional degeneracy: all coordinate
directions are radial directions. Due to the spatial isotropy of
the $a(\eta)_{x,y,z}$ functions, each
squared coordinate component of the gradient has the same value in the
local limit:
\begin{equation}
 (H_{x,y,z}^2)_0\,=\left(\frac{\dot a}{a}\right)^{\!\!2}_{\!0}.  
 \label{eq:hx02aa0}\end{equation}
It then follows from Equation~(\ref{eq:h2defini}) that
\begin{equation}
 H_0^2\,=3\,\left(\frac{\dot a}{a}\right)^{\!\!2}_{\!0}.  
    \label{eq:h023aa0}\end{equation}
The factor 3 in Equation \eqref{eq:h023aa0} 
appears because there are three mutually independent
coordinate labels (three degrees of freedom).

Note that the validity of Equations~(\ref{eq:hx02aa0}) and
  (\ref{eq:h023aa0}) is restricted to the local limit, as~marked by
  subscript 0. These equations are not valid beyond this limit, because~the
  symmetry of rotational invariance is broken. The~non-locality aspects
are addressed just next. 

So far, the parameters $H_{x,y,z}$, which are the components of a vector
field, and~$H^2$, which is a scalar, have only been defined in terms
of the spatial gradients of the scale factor $a$. None of those
quantities have yet been identified with the observable Hubble
parameter. As~soon as one wants to deal with an observable
like a redshift $z$, 
one has to select a viewing angle, a~line of sight to the object that
is being observed. The~line of sight relates the local ($r=0$)
observer with the non-local ($r>0)$ object. The~spatial isotropy is broken and the directional degeneracy is 
lifted as soon as one singles out a given line of sight to deal with a redshift
observation.

Let us examine the explicit effects of this by choosing the $x$
axis to be along the line
of sight. With~this choice, 
the $x$ coordinate equals the distance, i.e.,~$x=r$, while the
orthogonal coordinates  remain zero. Redshifts $z$ can only be
non-zero along lines of~sight.

When one forms the quadratic sum as in Equation~(\ref{eq:h2defini}) to
calculate the expression for $H^2$, only the contribution from the $x$
(line-of-sight) component has a redshift dependence. According to
Equations~(\ref{eq:hx2defin}), (\ref{eq:flrwaexpans}), and~(\ref{eq:a0overa}),
the $H_x^2$ contribution is obtained from the right-hand side of
Equation~(\ref{eq:flrwaexpans}) by replacing $a_0/a$ with the redshift
expression $1+z$. The~contributions $H_y^2$ and $H_z^2$ from the orthogonal directions must
be the same as $H_x^2$ in the local limit (where $r$ and the redshift $z$ are zero)
because of the isotropy requirement, but~there is no redshift dependence
in these two directions  because~the $y$ and $z$ coordinates remain
zero.

 Combining these three contributions, one obtains: 
\begin{equation}
H^2\,=\left(\frac{{\rm d}z}{{\rm d}r}
\right)^{\!\!2}=\frac{8\pi\, G}{3}\,\left[\,
\rho_{M_0}(1+z)^3\,+\,\rho_{R_0}(1+z)^4\,+\,2\,(\,
\rho_{M_0}\,+\,\rho_{R_0}\,)\,\right]. 
\label{eq:h2dzdrexpr}\end{equation}
The identification of $H$ as the redshift gradient ${\rm d}z/{\rm d}r$
in Equation \eqref{eq:h2dzdrexpr} 
follows from Equations~(\ref{eq:xlookback}) and (\ref{eq:zaeta0})
in combination with Equations~(\ref{eq:hx2defin}) and (\ref{eq:h2defini}).
The first two
$z$-dependent terms on the right-hand side come from the $H_x^2$
contribution along the line of sight. The~last term with a factor
of 2 comes from $H_y^2$ and $H_z^2$, which represent the two orthogonal 
directions with no $z$ dependence. Without~this constant term,
the requirement of spatial isotropy in the local limit is
violated. Equation~(\ref{eq:h023aa0}) is recovered 
from Equation~(\ref{eq:h2dzdrexpr}) \mbox{when $z=0$.}

$H^2$, as given in Equation~(\ref{eq:h2dzdrexpr}), is a scalar,
invariant with respect to spatial rotations and the choice of
coordinate system, in~contrast to the vector components
$H_{x,y,z}$. The~relation between the scalar $H^2$ and the vector
components $H_{x,y,z}$ is similar to the relation between the vector 
components $v_{x,y,z}$ of a velocity field and the kinetic energy,
which is a scalar quantity that is proportional to $v^2$.

While for convenience, the $x$ axis was
chosen to be along the line of sight in the present derivation, the~resulting
Equation~(\ref{eq:h2dzdrexpr}) is not dependent on this choice. Regardless of
the choice of coordinate system, the~equation can be described in a
coordinate-free way. There are three contributions to $H^2$. 
(a): The contribution along the line of sight, where the~spatial orientation is
determined by the viewing direction to the observed object. Only this
contribution 
has a redshift dependence. (b) and (c):  Two identical contributions from
the directions that are orthogonal to the line of sight. Although~the redshift
$z$ remains zero in these directions, their contributing terms are non-zero
and constant ($z$ independent) 
because of the boundary condition of spatial isotropy  for the scale factor gradient, which must be 
satisfied at the \mbox{$r=0$ boundary.} 

If the contributions from the two orthogonal directions is 
disregarded (as done in standard cosmology), the~Equation~\eqref{eq:h2dzdrexpr} for $H^2$ 
does not satisfy local rotational
invariance. Consistency with the FLRW assumption of spatial isotropy
requires the use of the isotropic boundary 
condition to allow us to correctly identify the 
scalar parameter $H(r)$ of Equation~(\ref{eq:h2dzdrexpr}) with the observable 
Hubble~parameter. 

It is standard praxis in cosmology to replace the
mass--energy density parameters $\rho$ with dimensionless parameters
$\Omega$. Equation~(\ref{eq:h2dzdrexpr}) can readily be converted into a more
convenient form:
\begin{equation}
H^2\,=H_0^2\,[\,
\Omega_M (1+z)^3 +\,\Omega_R(1+z)^4\,+ \Omega_X\,],
\label{eq:h2omegas}\end{equation}
where
\begin{equation}
H_0^2\,=\, 8\pi\, G\,(\,\rho_{M_0}\,+\,\rho_{R_0}\,)\,=\,
3\,\left(\frac{\dot a}{a}\right)^{\!\!2}_{\!\!0}
\label{eq:h02express}\end{equation}
as follows from Equations~(\ref{eq:flrwaexpans}) and
(\ref{eq:h023aa0}). Equation~(\ref{eq:h2omegas}) 
implies that the $\Omega$ parameters are normalized by the relation
\begin{equation}
\Omega_M + \Omega_R + \Omega_X =1.
\label{eq:normomegas}\end{equation}

Parameter $\Omega_X$ has been introduced to represent the constant
term that comes 
from the contributions to $H^2$ in the two directions that are
orthogonal to the line of sight. Its value,
\begin{equation}
\Omega_X\,=\, \frac{2}{3}\,,
\label{eq:omegax}\end{equation}
follows from Equations~(\ref{eq:h2dzdrexpr})--(\ref{eq:normomegas}).

Although the $\Omega_X$ term is mathematically identical to the
dimensionless cosmological constant $\Omega_\Lambda$ that is used by
standard cosmology, the~different index $X$ is used here to make it quite clear that this term is not a
cosmological constant. The~fundamental difference with respect
to standard cosmology is that their $\Omega_\Lambda$ term is also part
of the equation that governs the $a(t)$ function. In~contrast, 
Equation~(\ref{eq:flrwaexpans}) for $a(t)$ does not contain any such
constant~term. 

It follows from Equations~(\ref{eq:normomegas}) and 
(\ref{eq:omegax}) that
\begin{equation}
\Omega_M\,=\, \frac{1}{3}\,-\,\Omega_R.   
\label{eq:omegam}\end{equation}
Since $\Omega_R$, as~determined from the observed CMB (cosmic
microwave background) temperature,
 is about 0.0001, the~tiny deviation of $\Omega_M$ from 1/3 (assuming
spatial flatness) is not~measurable.

\section{Observational Evidence for the Non-Existence of Dark~Energy}\label{sec:nonexistde}
The treatment in Section \ref{sec:hubble} can be readily generalized to
the case when there is a cosmological constant $\Lambda$ in the
equation that governs the cosmic evolution. In~this generalized case,
Equation~(\ref{eq:flrwaexpans}) becomes 
\begin{equation}
\left(\frac{\dot a}{a}\right)^{\!\!2}\,=\frac{8\pi\, G}{3}\,\left[\,
\rho_{M_0}(a_0/a)^3\,+\,\rho_{R_0}(a_0/a)^4\,\right]\,+\,\frac{\Lambda}{3}. 
\label{eq:flrwaexpanslam}\end{equation}

The $\Lambda$-term 
propagates into Equations~(\ref{eq:h2dzdrexpr})--(\ref{eq:normomegas}) to give 
the following set of generalized equations: 
\begin{equation}
H^2\,=\frac{8\pi\, G}{3}\,\left[\,
\rho_{M_0}(1+z)^3+\rho_{R_0}(1+z)^4+\rho_\Lambda+\,2\,(\,
\rho_{M_0}+\rho_{R_0}+\rho_\Lambda\,)\,\right],
\label{eq:h2exprlam}\end{equation}
where
\begin{equation}
\rho_\Lambda =\frac{\Lambda}{8\pi\,G},
\label{eq:rholam}\end{equation}
while the version with the $\Omega$s becomes
\begin{equation}
 H^2\,=H_0^2\,\left[\,
\Omega_M (1+z)^3 +\,\Omega_R(1+z)^4\,+ \,\Omega_\Lambda +\,2/3\,\right],
\label{eq:h2omegaslam}\end{equation}
where
\begin{equation}
H_0^2\,=\, 8\pi\, G\,(\,\rho_{M_0}+\rho_{R_0}+\rho_\Lambda\,)\,=\,
3\,\left(\frac{\dot a}{a}\right)^{\!\!2}_{\!\!0}.
\label{eq:h02expresslam}\end{equation}
The generalized normalization condition for the $\Omega$s becomes
\begin{equation}
\Omega_M + \Omega_R + \Omega_\Lambda +2/3 =1.
\label{eq:normomegaslam}\end{equation}

Standard cosmology interprets the sum of the two constant terms as 
the cosmological constant,
\begin{equation}
\Omega_{\Lambda,\,{\rm stand.\,cosm.}}\,=\, \Omega_\Lambda +\,2/3,
\label{eq:ccstandcosm}\end{equation}
and determines its value by using it as a free parameter when fitting 
redshift data, as done by
\cite{stenflo-riessetal1998a,stenflo-perlmutteretal1999a} and in all
subsequent cosmological modeling  of the $z(r)$ relation. 
 According to Equation~(\ref{eq:ccstandcosm}),
however, one needs to 
subtract 2/3 from this fit to obtain the
value for the actual cosmological constant, which is relevant for the
dynamics and evolution of the universe. 
 Suitable examples among the numerous observational determinations that
have been reported in the literature are the ones from analysis of 
supernovae data by the Dark Energy Survey (DES) project, $\Omega_{\Lambda,\,{\rm
    stand.\,cosm.}} =0.669\pm 0.038$ \cite{stenflo-DESAbbott2019a} and
$0.648\pm 0.017$~\cite{stenflo-des2024a}, while Planck
Collaboration finds $\Omega_{\Lambda,\,{\rm
    stand.\,cosm.}} =0.685\pm 0.007$ \cite{stenflo-planck2020a}. These
and the various other values that have been reported seem to scatter
around the 2/3 value with a standard deviation of about
0.02. This leads to the conclusion that the current observationally
determined value for the cosmological constant, when
accounting for the required $\Omega_X =2/3$ term, is
\begin{equation}
\Omega_\Lambda \approx 0.00 \pm 0.02.
\label{eq:omlamobs}\end{equation}

This implies that there is no observational evidence for a non-zero cosmological
constant or for the existence of any ``dark energy''.
 Without a cosmological constant as a source, there is nothing else that can
 drive an acceleration expansion. Therefore, the 
expansion is slowing down, decelerating as expected from a
matter-dominated~universe. 

Since the value 2/3 for the constant term is independent of
the epoch in which the observer lives, there is no cosmic coincidence
problem, no violation of the Copernican principle. We are not
privileged~observers.

\section{Resolution of the \boldmath$H_0$ Tension}\label{sec:resh0tens}
The term ``$H_0$ tension`` refers to an inconsistency in
standard cosmology. The~derived value of the Hubble constant $H_0$ at the
present epoch depends on the types of data used. When it is based on
observations of redshifts and brightnesses of standard candles,
in particular of supernovae type Ia, one obtains $73.2\pm
1.3$~km~s$^{-1}$~Mpc$^{-1}$
\citep{stenflo-riessetal2018,stenflo-riess2021a}. This is about
9\%\ higher than the 
value of $67.4\pm 0.5~$km~s$^{-1}$~Mpc$^{-1}$ that is found from standard 
modeling of the CMB spectrum, which represents conditions in the
early universe \citep{stenflo-planck2020a}. While 9\%\ may not seem
like much, it is a 
4--5~$\sigma$ (standard deviation) 
discrepancy that has stubbornly resisted all elimination
attempts within the framework of standard~cosmology. 

The mentioned references represent seminal papers selected from very
extensive literature on the subject. For~a comprehensive set of
references, 
which include descriptions of alternative approaches to the
resolution of the $H_0$ tension, see 
the recent reviews in Refs.~\cite{stenflo-cervantes2023,stenflo-capozziello2024}. 

Even if the cosmological constant represents 
dark energy with an equation of state $-1$, it  could not be of significance for the physics of the early universe. The~$1/a^3$ and $1/a^4$ scalings of matter and radiation imply that 
$\rho_\Lambda \ll \rho_{M,R}$ when $a/a_0\ll 1$. Since $H_0$ is the
present value of the Hubble constant, it represents the local rather than the early~universe. 

The reason why $H_0$ has nevertheless been needed as a parameter in 
CMB modeling is that the Hubble radius $1/H_0$ provides a 
distance scale, which is part of the expression for the angular
diameter distance $D_\star$ to the last-scattering surface, where
matter and radiation decouple. The~relevant 
observed CMB quantity is the angular scale $\theta_\star$ of the
acoustic CMB peaks, which in the models is related to the
theoretically derived  radius $r_\star$ of the sound horizon at the epoch of
hydrogen recombination. The~defining relation is 
\begin{equation}\label{eq:thetastar}
\theta_\star\equiv 
{r_\star}/{D_\star},
\end{equation}
which through $D_\star$ depends on the value of $H_0$ for the local
universe. 

The angular diameter distance $D_\star=r_\star/\theta_\star$ is a well-determined model-independent parameter. The~angular anisotropy $\theta_\star$ is a directly observed 
quantity and is therefore independent of any model. $r_\star$ is
governed by known physics 
(assuming the validity of the standard model of particle physics) in
an early phase of the universe, when the cosmological constant did not
play any significant role. The~value of $D_\star$ that is derived from
the values of $\theta_\star$ and $r_\star$ therefore does not
depend on any model for the matter-dominated local universe. It is
only when one wants to use its established value to infer the value for the local Hubble
constant $H_0$ that the result becomes dependent on the model for the evolution of the scale factor in the local universe. This model depends on the nature and interpretation of the cosmological~constant. 

The meaning of the $D_\star$ parameter is to describe how the
linear $r_\star$ fluctuations on the surface of last scattering are 
mapped to angular anisotropies (represented by the angular scale parameter $\theta_\star$) of  the  locally measured radiation field: 
\begin{equation}\label{eq:dstara0a}
D_\star = \int_{t_\star}^{t_0}\frac{a_0}{a(t)}\,\,{\rm d}t,
\end{equation}
where $t_\star$ is the epoch of the last scattering surface. 
Using Equation~(\ref{eq:flrwaexpans}), one obtains: 
\begin{equation}\label{eq:dstardefy}
D_\star = \,\,\left(\frac{3}{8\pi\,G}\right)^{\!\!1/2}\!\!\int_{a_\star}^1
\,\frac{a_0\,{\rm
    d}a}{a^2\,[\,\rho_{M_0}(a_0/a)^3\,+\,\rho_{R_0}(a_0/a)^4\,]^{1/2}}\,\equiv
\,\,\,x_\star/H_0, 
\end{equation}
where $a_\star$ is the scale factor at the epoch of last scattering. 
$H_0=({\rm d}z/{\rm d}r)_0$ is the local value of the Hubble constant
that can be directly determined 
by redshift observations: it is given by
Equation~(\ref{eq:h02express}). The parameter $x_\star$ represents the angular diameter
distance $D_\star$ in units of the Hubble radius $1/H_0$.  In~explicit
form, $x_\star$  is given as
\begin{equation}\label{eq:xstarh0ey}
x_\star \,= \int_0^{z_\star} \!\!\frac{{\rm
    d}z}{[\,\Omega_M (1+z)^3 + \Omega_R (1+z)^4 \,]^{1/2}}.
\end{equation}
 
 The expression \eqref{eq:xstarh0ey} follows from the relation $8\pi\,G\,(\,
\rho_{M_0}+\rho_{R_0}\,)=3\,H_0^2\,(\,\Omega_M +\Omega_R\,)=3\,({\dot
  a}/a)_0^2$, which defines $H_0^2$ in the same way as in
Equation~(\ref{eq:h02express}) with the factor of 3, and~$\Omega_{M\,R}$ in the
same way as in Equations~(\ref{eq:h2omegas}) and (\ref{eq:omegam}),
with $\Omega_M +\Omega_R =1/3$. Note how the factor
of 3 in front of $H_0^2$ and the factor of 1/3 in $\Omega_M +\Omega_R$
compensate each other. The~meaning of $H_0$ and the definitions of
$\Omega_{M,\,R}$ are exactly the same as in
Section~\ref{sec:hubble}. Note that $x_\star$ in
Equation~(\ref{eq:xstarh0ey}) does not depend on 
any cosmological~constant. 

To make it explicit that the local value $H_0$ of the Hubble constant, 
which is defined in the last part of Equation~(\ref{eq:dstardefy}),
represents an observable local property of the 
$z(r)$ function that is obtained through redshift observations of
supernovae (SN) type Ia as standard candles, the subscript 
``SN'' is added to make the identification $H_0\equiv H_{0,\,{\rm SN}}$: this 
labeling distinguishes it from the value of $H_0$ that is found 
from modeling of CMB data. Equation~(\ref{eq:dstardefy})
then becomes
\begin{equation}\label{eq:h0sndef}
D_\star = x_\star/H_{0,\,{\rm  SN}}.
\end{equation}

The scale factor $a_\star$ is linked to the temperature $T_\star$
at which hydrogen recombination and decoupling between matter and
radiation occur, because~the temperature of the ambient radiation
field scales as $T\sim 1/a(t)$. Therefore, the value of $a_\star$ has
nothing to do with the observable redshift range but is obtained from
known recombination physics and direct $1/a$ scaling from the presently
measured CMB~temperature.

When standard cosmology (labeled ``sc'') calculates the angular diameter 
distance $D_{\star,\,{\rm sc}}$ from analysis of CMB data, the~starting point is the same, as shown in
Equation~(\ref{eq:dstara0a}). The~difference is that the $a(t)$ function, which
is used in that equation by standard cosmology, is governed
by a cosmological constant $\Omega_\Lambda$, and~that the Hubble
constant $H_0$ is assumed to represent the local expansion rate (${\dot
  a}/a)_0$. 

It is convenient to express $D_{\star,\,{\rm sc}}$ in the same form as
Equation~(\ref{eq:h0sndef}) by introducing the parameter $x_{\star,\,{\rm
  sc}}$ and by adding  the subscript 
 ``CMB'' to $H_0$ to make it explicit
 that it depends on CMB modeling with standard cosmology rather than on
redshift observations. Thus,
\begin{equation}\label{eq:dstarsc}
D_{\star,\,{\rm sc}} = \,\,x_{\star,\,{\rm sc}}/H_{0,\,{\rm CMB}}.
\end{equation}
Explicitly, 
\begin{equation}\label{eq:xstarsc}
x_{\star,\,{\rm sc}} \,= \int_0^{z_\star} \!\!\frac{{\rm
    d}z}{[\,\Omega_M (1+z)^3 + \Omega_R (1+z)^4 +
  \Omega_\Lambda\,]^{1/2}}.
\end{equation}
In contrast to Equation~(\ref{eq:xstarh0ey}), $x_{\star,\,{\rm sc}}$ depends on a 
cosmological constant, which affects the distance scale of the local universe. 

The values of $D_{\star,\,{\rm sc}}$ and $D_\star$ are identical
because they are both determined by the parameters $r_\star$ and $\theta_\star$
according to Equation~(\ref{eq:thetastar}), which do not depend on any
model for the local universe. Thus,
\begin{equation}\label{eq:dstardstarsc}
D_{\star,\,{\rm sc}} =D_\star=r_\star/\theta_\star.
\end{equation}
It therefore follows from Equations~(\ref{eq:h0sndef}),
(\ref{eq:dstarsc}), and~(\ref{eq:dstardstarsc}) that
\begin{equation}\label{eq:h0snratio}
\frac{H_{0,\,{\rm  SN}}}{H_{0,\,{\rm CMB}}} = \frac{x_\star}{x_{\star,\,{\rm sc}}}.
\end{equation}

Comparing Equations~(\ref{eq:xstarh0ey}) and (\ref{eq:xstarsc}), one
can see that $x_\star >x_{\star,\,{\rm sc}}$ because~the denominator in
Equation~(\ref{eq:xstarsc}) contains an extra $\Omega_\Lambda$ term,
while the $\Omega_{M,R}$ terms in the two equations are identical. 
After numerical evaluation of the $x_\star$ integrals with
$\Omega_\Lambda =2/3$, one finds
\begin{equation}\label{eq:h0ratiotheo}
\left(\frac{H_{0,\,{\rm SN}}}{H_{0,\,{\rm CMB}}}\right)_{\!\!\rm theory}\!\approx
1.093.
\end{equation}

This is to be compared with the observed value for the tension. According to the supernovae observations, $H_{0,\,{\rm SN}} =73.2\pm
1.3$\,km\,s$^{-1}$\,Mpc$^{-1}$ \cite{stenflo-riessetal2018,stenflo-riess2021a}  
, while CMB analysis in the framework of standard cosmology gives $H_{0,\,{\rm CMB}} =67.4\pm
0.5$~km~s$^{-1}$~Mpc$^{-1}$ \cite{stenflo-planck2020a}. 
The ratio between these two values represents the observed $H_0$ tension:
\begin{equation}\label{eq:h0ratioobs}
\left(\frac{H_{0,\,{\rm SN}}}{H_{0,\,{\rm CMB}}}\right)_{\!\!\rm obs}\!\approx
1.086\pm 0.021.
\end{equation}
The theoretical prediction is thus well within $1\sigma$ of the observed~value. 

Equation~(\ref{eq:h0snratio}) shows that if one removes the cosmological constant $\Omega_\Lambda$ from the $a(t)$
equation that is used by standard cosmology, then there is 
no $H_0$ tension at all, because~$x_{\star,\,{\rm sc}}$ becomes 
identical to $x_\star$.

\section{Age of the~Universe}\label{sec:age} 
The proper age of the universe is calculated in a way that is similar
to that for $D_\star$ in Section \ref{sec:resh0tens} with the identification
$H_0\equiv H_{0,\,{\rm SN}}$, with~the difference
that one has to use proper time instead of conformal time, and~that the
integration limit is the Big Bang instead of the surface of last
scattering. With~such modifications of Equations~(\ref{eq:dstardefy})
and (\ref{eq:xstarh0ey}), one obtains the following:
\begin{equation}\label{eq:tage}
t_0 =\int_0^{t_0}\!{\rm d}t =\,\,\frac{1}{H_{0,\,{\rm SN}}}\,\int_0^\infty \!\!\frac{{\rm d}z}{(1+z)\, [\,\Omega_M(1+z)^3 +\Omega_R(1+z)^4\,]^{1/2}}.
\end{equation}
As always, throughout the present paper, $\Omega_M +\Omega_R =1/3$
according to Equation~(\ref{eq:omegam}). 

Since $\Omega_R\ll \Omega_M$, the age $t_0$ can be obtained more
directly and given quite a simple analytical form. Numerical evaluation of
the full integral when $\Omega_R$ is accounted for turns out to give a
value for $t_0$ that differs by less than one per mille from the value
obtained when $\Omega_R$ is set to zero. Therefore, one  
can treat the universe as entirely matter-dominated. 
As $a\sim t^{2/3}$ in this case, the age is 2/3 of the linearly
extrapolated age $(a/{\dot a})_0$.
 Since $({\dot a}/a)_0 =H_0/\sqrt{3}$ according to
Equation~(\ref{eq:h02express}), one obtains the following:
\begin{equation}\label{eq:tageanalyt}
t_0 = \frac{2}{H_{0,\,{\rm SN}}\, \sqrt{3}}.
\end{equation}

The result \eqref{eq:tageanalyt} 
can also be obtained via direct analytic integration of
Equation~(\ref{eq:tage}) with \mbox{$\Omega_M =1/3$} and $\Omega_R =0$.

Using the supernovae value of 73.2~km~s$^{-1}$~Mpc$^{-1}$ for the
Hubble constant, one finds $t_0 =15.4$~Gyr. 
This is to be compared with the value $t_{0,\,{\rm sc}} =13.8$~Gyr
that has been adopted by  standard cosmology. It is based on the expression 
\begin{equation}\label{eq:tagesc}
t_{0,\,{\rm sc}} =\,\,\frac{1}{H_{0,\,{\rm CMB}}}\int_0^\infty
\!\!\frac{{\rm d}z}{(1+z)\,[\,\Omega_M(1+z)^3 +\Omega_R(1+z)^4
  +\Omega_\Lambda\,]^{1/2}},
\end{equation}
which is similar to Equation~(\ref{eq:tage}), with~the difference that
it contains a cosmological constant and~that the values that are
used, $H_{0,\,{\rm CMB}} = 67.4$\,km~s$^{-1}$~Mpc$^{-1}$ and
$\Omega_\Lambda =0.685$, have been inferred from the Planck mission
CMB data~\cite{stenflo-planck2020a} within the interpretational
framework of standard~cosmology.

\subsection{Look-Back~Times}\label{sec:lookback} 
The look-back time $t_{\rm lb}$ to a galaxy with redshift $z_{\rm lb}$
is obtained from the expressions for the proper age $t_0$ of the
universe by replacing the upper integration limit $\infty$ with
$z_{\rm lb}$. A~convenient way to see how the look-back times $t_{\rm lb}$ of the
present theory are related to those of standard cosmology is to
introduce a function $r(z)$, defined as 
\begin{equation}\label{eq:tlbrz}
t_{\rm lb} \equiv\,\,\frac{1}{H_{0,\,{\rm CMB}}}\int_0^{z_{\rm lb}}
\!\!\frac{r(z)\,\,{\rm d}z}{(1+z)\,[\,\Omega_M(1+z)^3 +\Omega_R(1+z)^4
  +\Omega_\Lambda\,]^{1/2}}.
\end{equation}
Comparison with
Equation~(\ref{eq:tagesc}) shows that if the function $r(z)$ is replaced by unity,
then one obtains the look-back times of standard cosmology. When 
$r(z)$ is instead
\begin{equation}\label{eq:rzexpl}
r(z) \,=\,\,\frac{H_{0,\,{\rm CMB}}}{H_{0,\,{\rm
      SN}}}\,\left[\,1\,+\,\frac{2}{(1+z)^3}\,\right]^{1/2}, 
\end{equation}
comparison with Equation~(\ref{eq:tage}) shows that one obtains
the look-back times of the present theory. To~obtain the 
form of
Equation~(\ref{eq:rzexpl}), the radiation contribution $\Omega_R$ has
(as before) been disregarded because~it is $\ll$$\Omega_M$. This allows
us to express $\Omega_\Lambda/\Omega_M$ as~2. 

The right-hand side of Equation~(\ref{eq:rzexpl}) consists of two
competing factors. The~first factor, $H_{0,\,{\rm CMB}}/H_{0,\,{\rm
    SN}}$, is redshift-independent and equals 0.92 according to
Equation~(\ref{eq:h0ratioobs}). It is due to the $H_0$ tension. While
this factor reduces the 
look-back times, the~second, $z$-dependent factor has the opposite
effect, enhancing the look-back times. This enhancement increases with
decreasing redshift 
and equals $\sqrt{3}\approx 1.73$ in the
local limit where $z=0$. When the redshift goes to infinity
the factor goes to~unity.

The net effect of these two competing effects is that the look-back
time to the cosmic horizon (the Big Bang), which represents the proper age of
the universe, is $15.4/13.8\approx 1.12$ times larger in the present
theory compared with standard~cosmology.

\subsection{Comparison with Stellar~Ages}\label{sec:stellar} 
First-generation (population~III ) stars have not yet been identified,
so the oldest known stars are low-metallicity second-generation
population~II stars 
in the halo of our galaxy, often as members of globular clusters
(GC). While all stars must be younger than the age $t_0$ of the
universe, one 
can better constrain the admissible stellar ages by
accounting for the time needed to form the first stars. Numerical
simulations suggest that the rate of GC formation and the halo of the
Milky Way peaked about 
0.4--0.6~Gyr after the Big Bang
~\cite{stenflo-trenti_etal2015,stenflo-naoz2006}. This leads to 
adopt ($t_0 -0.4$)~Gyr as a realistic upper limit for the age of any observable~star. 

\begin{figure}[H]
 \includegraphics[width=0.7\columnwidth]{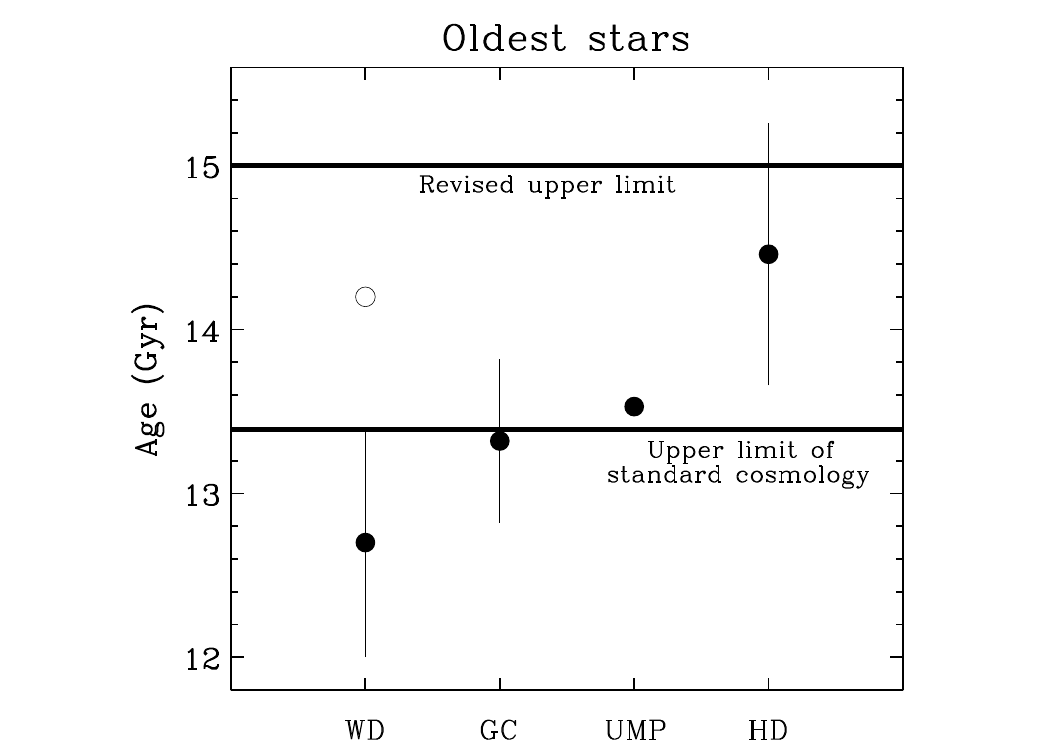}
\caption{Using the oldest stars to 
  discriminate between cosmological models. The~upper age limit of
  observed stars
  is defined as ($t_0 -0.4$)~Gyr, the~estimated age of
  the Milky Way halo (thick horizontal lines). 
Four stellar age determinations are selected, denoted as WD, GC, UMP, and~HD.
 See text for details. 
}\label{fig:oldstar}
\end{figure}

With this choice, the upper
stellar age limit is ($13.8-0.4)$~Gyr~$=13.4$~Gyr according to standard
cosmology, while it is $(15.4-0.4)$~Gyr~$=15.0$~Gyr with the present
theory. These two upper limits are  marked by thick horizontal lines
in Figure~\ref{fig:oldstar}.

  There is some uncertainty in choosing these limits. They are significantly lowered and more 
  constraining with the choice ($t_0 
  -0.6$)~Gyr. Therefore ($t_0 -0.4$)~Gyr  is a more conservative~choice.

To highlight the role of accurately determined stellar ages for the
discrimination between cosmological theories,
  a few prominent stellar cases are discussed here.
Thus, four stellar cases have been selected in 
Figure~\ref{fig:oldstar}, which 
have been particularly well studied
using a variety of age determination methods.

The star denoted ``HD'' is the so-called ``Methuselah'' star HD
140283, a~population~II halo subgiant at distance 
202~ly with a metallicity 250 times smaller than that of the
Sun. Modeling with
isochrones gives the age ($14.46\pm 0.8$)~Gyr~\cite{stenflo-bond2013a}, which 
lies somewhat  beyond the range of standard~cosmology. 

The star denoted ``UMP'' reresents the ultra metal-poor (UMP) low-mass binary star 
system 2MASS J18082002–5104378. Spectroscopic analysis and isochrone
fitting give a reported age of ($13.53\pm 0.002$)~Gyr
~\cite{stenflo-schlaufman2018}. No estimates of the systematic
uncertainties are available for this star. . 

Since the UMP binary system has a Galactic thin disk orbit, it has
been suggested~\cite{stenflo-schlaufman2018} that the $13.5$~Gyr
age represents a lower 
limit for the age of the thin Galactic disk. This contradicts the
value of ($8.8\pm 1.7$)~Gyr for the age of the thin disk that has been determined 
using Th/Eu
nucleocosmochronology~\cite{stenflo-peloso2005}. The~contradiction may
 be resolved if the binary system were formed long before the formation of the~disk. 

The system denoted ``GC'' represents a set of 68 globular clusters
~\cite{stenflo-valcin_etal2020}. The~age that is quoted, ($13.32\pm 
0.5$)~Gyr, is an average over the 38 clusters that have
metallicity [Fe/H]~$<-1.5$ after~applying an ad~hoc prior that
excludes values larger than 15~Gyr. However, since there is an age
distribution 
within this subset that exceeds the
measurement uncertainties, the~average value is not representative of
the upper age limit of the clusters. The actual globular cluster age upper
limit is therefore expected
to lie significantly higher than the plotted
point. Different viewpoints on the physics of 
GC formation~\cite{stenflo-forbes_etal2018,stenflo-krumholz_etal2019} 
are consistent with significantly increased GC age~limits.  

Finally, ``WD'' denotes white dwarfs (WDs) for which the age
is determined from white dwarf cooling. The~observed
luminosity function of white dwarfs in the globular cluster M4 was 
fitted with theoretical models to give an age of ($12.7\pm 0.7$)~Gyr
~\cite{stenflo-hansen2002a}. The~same model 
parameters were used for white dwarfs in the Galactic disk, giving 
($7.3\pm 1.5$)~Gyr. This is 1.5~Gyr lower than
the mentioned 8.8~Gyr value determined from 
nucleocosmochronology~\cite{stenflo-peloso2005}. While the absolute
age scale of the white 
dwarf cooling method is sensitive to the model parameters,
the~relative age difference of ($12.7-7.3$)~Gyr~$=5.4$~Gyr between the
halo and the disk is not. If~one raises the disk 
age  from 7.3~Gyr  to 8.8~Gyr to make it consistent with
nucleocosmochronology while retaining consistency of the WD cooling
method, then 
the halo age needs to be raised from 12.7~Gyr to 
14.2~Gyr; this value is plotted 
 as an open circle in
Figure~\ref{fig:oldstar}; the value plotted 
 lies well beyond the upper age limit of standard cosmology. 

Asteroseismology has considerable potential as a major and independent tool for the determination of stellar ages. The~CoRoT and Kepler satellites have recorded oscillating
modes for many thousands of stars across the HRD (Hertzsprung--Russel diagram)  ~\cite{stenflo-silvaaguirre2016}. This database has been used to 
determine the vertical age gradient for the red giant stars in the Galactic disk~\cite{stenflo-casagrande2016a}. The~derived stellar age distribution has a tail that extends well beyond 14~Gyr, although~it
is not yet clear enough 
whether this tail is an artifact of the analysis or whether it is in
conflict with the assumed standard age of the universe. 
One expects the uncertainties to come down sufficiently in the~future.

\section{Conclusions}\label{sec:concl}
The evolution of the universe on cosmic time scales can only be
observed by examining the static sequence of historical events along each
line of sight when one looks out in space and back in time. Distance and
look-back time are equivalent coordinates along light rays. Studying them is like
conducting cosmic archaeology.
The~observable function is $z(r)$, redshift versus distance. The~theoretical,
dynamic description is scale factor versus time,
the~$a(t)$ function. To~make contact between theory and observations,
one must know the 
relation between the $a(t)$ and $z(r)$ functions.

In the past, this has seemed like a trivial problem. Standard cosmology
has just used the relation $1+z=1/a$ to replace the $a$-dependent terms in
the equation that governs the $a(t)$ function with $z$-dependent
terms and~identified the expansion rate ${\dot a}/a$ with the Hubble
constant $H={\rm d}z/{\rm d}r$ that describes the distance dependence
of the~redshift. 

This procedure has overlooked the profound geometric difference between dynamic,
proper time $t$ and look-back time $t_{\rm lb}$ along lines of
sight. In~contrast to $a(t)$, which is a 1D function, $a(t_{\rm lb})$ is an isotropic
function in a 3D space with an observer at its center. While
the requirement of isotropy must be satisfied locally around each
spacetime point, the~symmetry is broken when an observation is performed
because a particular line of sight is singled out, which connects the
observer with the observed object. In~the
local limit, however, the~symmetry of isotropy is not broken, there
is directional degeneracy, and any direction is 
a radial direction. To~satisfy the requirement of spatial isotropy in
the local limit, the~spatial gradients of $a$ in  all
three coordinate directions must be accounted~for. 

Due to the local isotropy requirement, the gradients in the two
directions that are orthogonal to the chosen line of sight give rise to a
constant term in the equation that governs the $z(r)$ function. This
term, denoted here as $\Omega_X$, mathematically plays the same role as
the cosmological 
constant $\Omega_\Lambda$ in standard cosmology, but~it is neither a free
parameter nor a cosmological constant. It has the same value, 2/3, for~any observer at any epoch in cosmic 
history, and~it does not appear in the equation for $a(t)$.

The 2/3 value agrees within the observational errors with the
value that is found for the $\Omega_X$ term when it is used as a free parameter
in model fits of the observational $z(r)$ function that represents
the observed redshifts of supernovae type Ia as standard candles.
A hypothetical non-zero cosmological constant adds
to the 2/3 value for the constant term. Since there is no observational evidence that the
constant term significantly differs from 2/3, there is no evidence in
support of the existence of ``dark~energy''.

Our universe is therefore decelerating, slowing down as predicted by
its contents of matter and radiation. Since the 2/3 value is independent of
the epoch in which the observer lives, there is no ``cosmic
coincidence''~problem. 

In standard cosmology, both the $a(t)$ and the $z(r)$ functions are
governed by the same cosmological constant $\Omega_\Lambda$. In~the
present theory, $\Omega_X =2/3$ is only part of the equation for $z(r)$,
but not of the equation for $a(t)$. This difference has observational
consequences. Here, two of them have been discussed, the~$H_0$ tension and
the age of the~universe.

With the present theory, the age of the universe is increased 
from 13.8~Gyr to 15.4~Gyr. This relieves some
tension with the determined stellar ages for some of the oldest known
stars. The~error bars of stellar ages, however, need to be reduced
before this difference can be used as a suitable cosmological~test. 

An observational validation of the theory is, however,
already available: The theory resolves the $H_0$ tension and
predicts its observed magnitude without the use of any free 
parameters. When the cosmological constant is removed from
the equation for the $a(t)$ function that is used by standard
cosmology in the CMB
analysis, the~systematic difference
between the value of the Hubble constant $H_0$ determined from the
redshifts of supernovae and the value determined from
modeling of CMD data disappears. Therefore, consistency of cosmological theory
is~restored.

While this kind of explanation of the cosmological constant and the
$H_0$ tension has been given by the present author before 
(see~\cite{stenflo-2023}, which also contains references to earlier
versions of the theory), the~difference is that the previous versions
depend on a perturbation approach and arguments from quantum
physics. Thus, possible boundary
conditions for metric fluctuations constrained within the finite
time-frozen 3D subspace that represents the observable universe were
explored. The~requirement that there should be no observable
gravitational effects from such metric vacuum fluctuations was
shown to be satisfied by a discrete set of fluctuation modes. Each
mode corresponds to a boundary term that formally appears as a
cosmological constant. If~one makes use of arguments from quantum
physics that the universe should be in the mode with the lowest allowed
energy state, then a unique value for the constant $\Lambda$-type
boundary term is obtained. It is found to agree with the
observationally determined value for the cosmological constant, without~the use of any free parameters in the~theory.  
 
In contrast, the present version of the theory does not make use of any
perturbation analysis or heuristic arguments from quantum physics. The~derivation is entirely~classical.

\vspace{6pt} 

\funding{This research received no external~funding.}

\dataavailability{No new data were created or analyzed in this
  study. Data sharing is not applicable to this article.}

\acknowledgments{The author expresses gratitude to \AA ke Nordlund for many helpful~discussions.}

\conflictsofinterest{The author declares no conflicts of~interest.}

\begin{adjustwidth}{-\extralength}{0cm}

\reftitle{References}


\PublishersNote{}

\end{adjustwidth}
\end{document}